\documentclass[12pt]{iopart}

\usepackage{graphicx}
\usepackage{iopams}
\usepackage{color}

\begin{document}


\title[Dynamical Phase Transition in a Two-Species Eden Model]{Range Expansion with Mutation and Selection: \\ Dynamical Phase Transition in a Two-Species Eden Model}

\author{J.-T.~Kuhr, M.~Leisner, and E.~Frey\footnote{Author to whom any correspondence should be addressed.}}

\address{Arnold Sommerfeld Center for Theoretical Physics (ASC) and Center for NanoScience (CeNS), Ludwig-Maximilians-Universit\"at M\"unchen, Theresienstra\ss e 37, 80333 M\"unchen, Germany
}%
\ead{frey@lmu.de}
\date{\today}

\begin{abstract}
The colonization of unoccupied territory by invading species, known as range expansion, is a spatially heterogeneous non-equilibrium growth process. We introduce a two-species Eden growth model to analyze the interplay between uni-directional (irreversible) mutations and selection at the expanding front. While the evolutionary dynamics leads to coalescence of both wild-type and mutant clusters, the non-homogeneous advance of the colony results in a rough front. We show that roughening and domain dynamics are strongly coupled, resulting in qualitatively altered bulk and front properties. For beneficial mutations the front is quickly taken over by mutants and growth proceeds Eden-like. In contrast, if mutants grow slower than wild-types, there is an antagonism between selection pressure against mutants and growth by merging of mutant domains with an ensuing absorbing state phase transition to an all-mutant front. We find that surface roughening has a marked effect on the critical properties of the absorbing state phase transition. While reference models, which keep the expanding front flat, exhibit directed percolation critical behavior, the exponents of the two-species Eden model strongly deviate from it. In turn, the mutation-selection process induces an increased surface roughness with exponents distinct from that of the classical Eden model.
\end{abstract}



\maketitle

\section{\label{sec:intro}Introduction}

Spatial structure strongly influences the dynamics of evolving biological systems and often gives rise to qualitatively different outcomes as compared to well-mixed populations~\cite{Durrett:1994ua,murray2007book2}. During the last years microbiological experiments have progressively been used to shed light on the dynamics of spatially extended systems~\cite{Shapiro:1995id,Jacob:1999bn,Kerr:2002kk,BenJacob:2003kn,Buckling:2009je,Beer:2010iu,Zhang:2010jn}, and spurred theoretical investigation~\cite{Reichenbach:2007hk, Reichenbach:2008kf, Wakano:2003ip, Cates:2010hx, Wakano:2009hy, Wakano:2011fs, Melke:2010ij}. Since microbial colonies typically grow from some initial seed, a particularly interesting question to ask is how both a colony's morphology \emph{and} its internal composition are shaped by the growth rates of the different strains it is composed of, and by the interactions between these strains. Already the simplest scenario, the spreading of two selectively neutral strains or species shows intriguing phenomena which makes the evolutionary outcome quite distinct from well-mixed populations~\cite{Klopfstein:2006bl, Hallatschek:2007gv, Excoffier:2008hw, Korolev:2010cz, Ali:2010do, Hallatschek:2010kb}. Experimental investigations of expanding \emph{E.~coli} and \emph{S.~cerevisiae} colonies containing two fluorescently labeled but selectively neutral strains have shown that the population differentiates along the growing front and thereby segregates into well-defined domains~\cite{Hallatschek:2007gv,Hallatschek:2010kb}. This is caused by demographic fluctuations: Since mainly cells at the leading front of the growing colony access nutrients and reproduce, the effective population size is small and neutral dynamics leads to local fixation of strains and thereby generates sectoring of the population~\cite{Hallatschek:2010kb}. 

In general, however, microbial communities are heterogeneous and composed of multiple strains, which may have different growth rates or show other kinds of distinct phenotypic features, cf.~e.g.~\cite{Alberti:1990wf, Melke:2010ij, Nadell:2009de, Griffin:2004kn, Wilson:2010us, Tuchscherr:2011fz, Neulinger:2008gv}. One well-known phenomenon, which is the focus of this work, are bursts of new sectors of mutants during the growth of bacterial colonies~\cite{Koshland:1985tg, Shapiro:1995id, BenJacob:1998ff, Golding:1999wa, Hallatschek:2010kb}. In a growing bacterial colony, which initially consists of one phenotype (the wild-type) only, mutations of this strain may appear during reproduction of individuals. If these mutations happen at the leading front and are beneficial, i.e., if the mutant strain has a larger growth rate than the wild-type strain, mutant regions along the front not only advance faster but also expand laterally. Consequently mutant sectors take over ever larger parts of the front in a quasi-deterministic fashion~\cite{Saito:1995wl,Murray:2007ux,Hallatschek:2010kb}. While deleterious mutations are, for the same reason, handicapped by selection, they may still form large clusters along the front, if they appear frequently enough. In this case mutant sectors are no longer spatially separated, but may coalesce. At a critical mutation probability, the mutants' selective disadvantage is effectively balanced and mutants may take over the front~\cite{Hallatschek:2010kb}. If back-mutations are prohibited, the front remains trapped in an all-mutant state. In the language of non-equilibrium statistical mechanics, the critical mutation rate marks a phase transition between an active state, for which the front is composed of wild-types and mutants, and an absorbing homogeneous state, composed of mutants only~\cite{Hinrichsen:2000wg, Hinrichsen:2006cn, Odor:2004wm, henkel2009book1}. 

What makes this absorbing state phase transition in a growing bacterial colony interesting is the intricate interplay between the morphology of the growing front and the evolutionary dynamics of the colony. Depending, among others, on the particular type of bacterial strain, nutrient concentration and softness of the agar surface, bacterial colonies exhibit a kaleidoscope of possible morphologies~\cite{Budrene:1991gq, Shapiro:1995id, BenJacob:1998ff, Jacob:1999bn, BenJacob:2003kn, Matsushita:1999vr, Matsuyama:2001vy}. The Eden model~\cite{Eden:1960vd} has been devised to describe the growth of bacterial cultures with a compact morphology on which we focus in this work. A hallmark of this model is the generation of rough fronts with characteristic features closely resembling recent experimental observations~\cite{Huergo:2010bb}. The roughness of the front directly affects the trajectory of interfaces between domains of different strains~\cite{Saito:1995wl}: the undulations of the front are imprinted in the meandering of the domain boundaries on all length scales, as has recently also been observed experimentally~\cite{Hallatschek:2007gv,Hallatschek:2010kb}. This surface-induced meandering speeds up coalescence of clusters, i.e., the roughness of the propagating front strongly affects the temporal evolution of the population's composition. This suggests that front roughening is highly relevant for the nature of the phase transition from a heterogeneous to a homogeneous population at the expanding front. 

It is precisely this issue which we would like to address in this manuscript.  To this end we study bacterial range expansion using a two-species Eden model, which incorporates surface roughness, selection and irreversible mutations. We intend to gain deeper insight into the interplay of these key features of the dynamics and their relative importance for the transition to the absorbing state. While our findings are mainly of general importance for a broader class of multi-species growth models, we also expect that real world range expansions of bacterial colonies are subject to this coupling and carry its signature in the evolving patterns. In the remainder of the introduction we give a concise overview of previous work on surface roughness and absorbing state phase transitions as relevant for this work. A more in-depth discussion and comparison with the results of our work is given in the final section.

Both, discrete numbers and roughness of the expanding front, are intrinsic to surface growth models~\cite{godreche1991solids,HalpinHealy:1995wb,Barabasi:1995p4091}, which mimic the stochastic advance of particles into empty space. One particular growth model, the Eden model~\cite{Eden:1960vd} (of which there are three, slightly different, variants~\cite{Jullien:1985wv}) has been devised to describe the growth of bacterial cultures. Mesoscopically its evolution is captured by the Kardar-Parisi-Zhang (KPZ) equation~\cite{Kardar:1986vl}. The KPZ equation constitutes a robust universality class which incorporates many surface growth models, like e.g.~ballistic deposition and solid-on-solid models. There have been a number of generalizations to multi-species growth models~\cite{Derrida:1991tv, Saito:1995wl, Ausloos:1993ts, Ausloos:1996vs, Pellegrini:1990uk, Wang:1993uq, Wang:1995wj, ElNashar:1996ts, ElNashar:1998nh, Kotrla:1997vx, Kotrla:1998vf, Drossel:2000vw, Drossel:2003cwa, Reis:2002jn}, notably one of the Eden model which incorporates selection~\cite{Saito:1995wl}. The coupled influence of mutations and selection on kinetic surface roughening, which is one of the topics of this work, has not been analyzed in detail so far. Even when neglecting roughness, multi-species propagation is not treated easily in more than one dimension. The reason for this is the intricate interplay of creation, annihilation and merging of clusters, which contain only one kind of individuals, at the leading front of the colony. 

In the case of irreversible mutations both analytical results and numerical simulations for range expansion with flat fronts (neglecting surface roughness) predict a transition to an all-mutant absorbing state, which emerges even for deleterious mutations at a critical mutation probability~\cite{Hallatschek:2010kb}. For such models with flat fronts~\cite{Hallatschek:2010kb}, the dynamics closely resembles that of  a contact process~\cite{Harris:1974ue}. As a consequence, it belongs to a broader class of absorbing state phase transitions whose main representative is directed percolation (DP)~\cite{Broadbent:2008fr}, a dynamic version of percolation~\cite{stauffer1994book}. The DP universality class of phase transitions to absorbing states has been found to display an enormous robustness with respect to alterations of the microscopic update rules and is considered a paradigm of non-equilibrium statistical physics~\cite{Hinrichsen:2000wg, Hinrichsen:2006cn, Odor:2004wm, henkel2009book1}. How roughness of the front may influence phase transitions to absorbing states has, to the best of our knowledge, previously not been addressed.

The paper is organized as follows. In Section \ref{sec:model} we introduce a two-species growth process on a 2$d$ lattice to analyze the general properties of range expansions of asexually reproducing microorganisms. It explicitly includes irreversible mutations from wild-types to mutants and selection between the two strains. The evolving system's morphology intimately depends on the antagonistic effects of new mutant domains being created at the front and others loosing contact to it. 
For abundant, deleterious mutations a  phase transition to an absorbing state exists, which changes both the evolutionary dynamics and surface roughening behavior of the system qualitatively. We discuss properties of both surface and bulk morphology and map out the phase diagram in Section \ref{sec:pandp}. The system's critical behavior near the transition is affected by the front's roughness, since the temporal evolution of the system is restricted to the  growing front. This alters the critical properties of the absorbing state phase transition. In Section \ref{sec:bulk} we determine its critical exponents, which are different from those of the DP class (which is most often found for flat systems~\cite{Hinrichsen:2000wg, Hinrichsen:2006cn, Odor:2004wm, henkel2009book1}). In addition, the different birth rates of the two strains induce an enhanced width of the front near the phase transition. Close to the transition the roughness exponent of our model is severely enhanced compared to that of the Eden model~\cite{godreche1991solids,HalpinHealy:1995wb,Barabasi:1995p4091}. In addition to KPZ behavior we identify and characterize a critical roughening regime in Section \ref{sec:surface}. We conclude 
with a discussion of our results and a comparison to related models which study the coupling between surface roughening and domain dynamics.

\section{\label{sec:model}Eden Model with Mutations}

Range expansion into hitherto unoccupied territory proceeds in a non-homogeneous manner on the length scale of individuals. Along the leading front local protrusions emerge randomly and subsequently expand, thereby creating a rough front and an overall forward movement. The main features of this growth process are well captured by the classical Eden model, which was developed to mimic the growth of microbial colonies~\cite{Eden:1960vd,HalpinHealy:1995wb,Barabasi:1995p4091}. 
While some multi-species extensions of the Eden model have been analyzed~\cite{Derrida:1991tv,Saito:1995wl,Ausloos:1993ts, Ausloos:1996vs}, surface growth experiments with competing  microorganisms have only been performed recently~\cite{Hallatschek:2007gv,Hallatschek:2010kb}. They reveal intriguing, non-trivial patterns if the population is comprised of distinguishable sub-populations. Mutations and selection can alter the growth dynamics by giving some individuals a growth advantage or by introducing qualitatively different organisms. 

\begin{figure}[htb]
\centering
\includegraphics{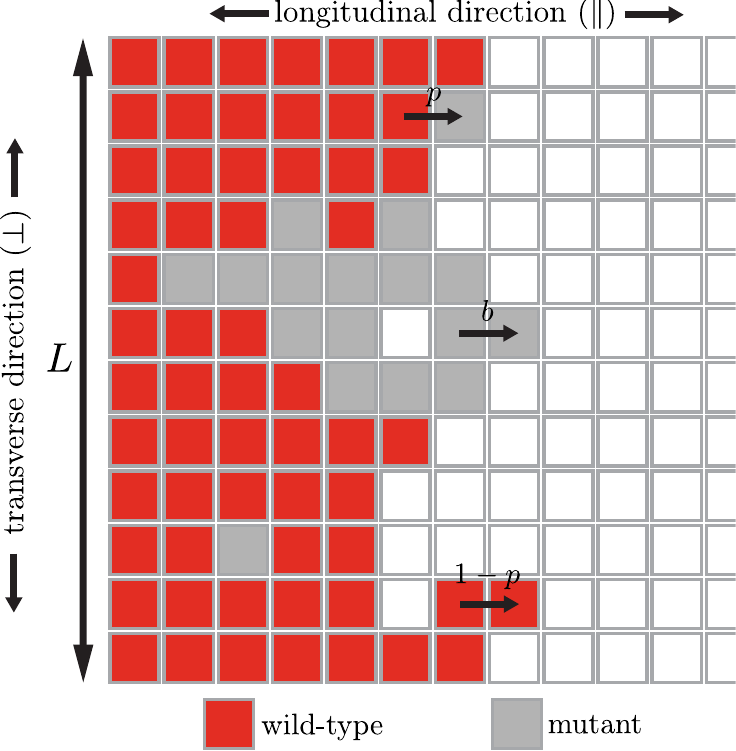}
\caption{\label{fig:model} Eden model with mutations (color online). An initially line-shaped bacterial colony of length $L$, consisting of only wild-types, shown in dark gray (red), grows into an empty half-space. Wild-types reproduce at rate $1$, given that they have a vacant nearest-neighboring lattice site. Individual offsprings are either wild-types or mutants (light gray) with probabilities $1-p$ and $p$, respectively. Mutants reproduce like wild-types but at birth rate $b$. Back-mutations are prohibited. In the $L$-direction (transverse) periodic boundary conditions apply.}
\end{figure}

In this work we employ a lattice gas model to analyze the influence of mutations and selection on range expansion at rough, fluctuating fronts. We model microbial range expansion with mutations as a cellular automaton on a 2$d$ semi-infinite square lattice of extensions $L \times \infty$ with periodic boundary conditions in the transverse direction (see Fig.~\ref{fig:model}). 
At a given time $t$, each site $(i,j)$, with $i \in \{1,\ldots L\}$ and $j \in \mathbb{N}$, is either empty or occupied with an individual, which in turn can be either a wild-type or a mutant. We identify empty sites, wild-types, and mutants with state variables $s=0$, $s=1$, and $s=-1$, respectively. The state of the system at a given time is, therefore, specified by the set of occupation numbers $s_{i,j} \in \{-1,0,1\}$. The system evolves as individuals reproduce: empty sites with $s = 0$ can change their state if an individual on a nearest-neighboring site reproduces. We assume that individuals do not die and hence any site with $s = \pm 1$ remains in its state indefinitely.

To implement random-sequential update of a configuration we apply a simplified version of Gillespie's stochastic algorithm~\cite{Gillespie:1976vd}. Only individuals at the front, defined as the set of occupied sites with at least one empty nearest-neighbor site, can reproduce. Of these, an individual is randomly, but proportional to its birth rate, chosen to reproduce: Mutants reproduce with relative birth rate $b \ge 0$, while the birth rate of wild-types is 1 and thereby sets the timescale. Let $N_{wt}$ and $N_{mut}$ denote the number of wild-types and mutants with free neighbors, respectively. The overall birth rate of the population (at which production events happen) is given by $b_{tot} := (N_{wt} + bN_{mut})$. To account for different birth rates, each wild-type individual with an empty neighbor is chosen to reproduce with probability $1/b_{tot}$, while each mutant individual with an empty neighbor is chosen with probability $b/b_{tot}$. The new individual is placed on a random empty nearest-neighbor site of the individual which has been chosen to reproduce. During wild-type reproduction, a mutation may happen with probability $p$. Thus, if the reproducing individual is wild-type ($s = 1$), the offspring is a wild-type with probability $1-p$ and a mutant with probability $p$. If the reproducing individual is a mutant ($s = -1$), the offspring is necessarily also a mutant. Note, that by these rules  back-mutations and multiple mutations are prohibited, and that all mutants are identical and reproduce with the same birth rate $b$. Assuming exponentially distributed reproduction times, the expectation time until the next reproduction event is $b_{tot}^{-1}$. Hence, we update time by $t \rightarrow t + b_{tot}^{-1}$ whenever an individual reproduces. Since, after some initial transient period, the average front position moves at constant velocity, one may take the longitudinal coordinate $j$ as a proxy for time $t$.

The model, as described above, is a generalization of version C of the Eden model as introduced by Jullien and Botet~\cite{Jullien:1985wv}. It is the biologically most realistic version, as it focusses on occupied sites, i.e., individuals, rather than on empty sites (version A) or bonds between adjacent occupied and empty sites (version B). In the limit of vanishing mutation rate our model reduces to the model of Saito and M\"uller-Krumbhaar~\cite{Saito:1995wl}. While we consider the case of a homogeneous initial front with uni-directional mutations, they analyzed the temporal evolution of an initially heterogeneous front in the absence of mutations, see Section \ref{sec:discussion} for more details. If not stated otherwise, we use a line of wild-type particles as initial condition, i.e., $s_{i,j}= \delta_{j,1}$. Since diffusion is not included in the model, surface configurations become frozen in the bulk, as observed for patterns in range expansion experiments~\cite{Hallatschek:2007gv}. 

\section{\label{sec:pandp}Phenomenology and Phase Diagram}
\subsection{\label{sec:phenomenology}Phenomenology }

\begin{figure}[thb]
\centering
\includegraphics{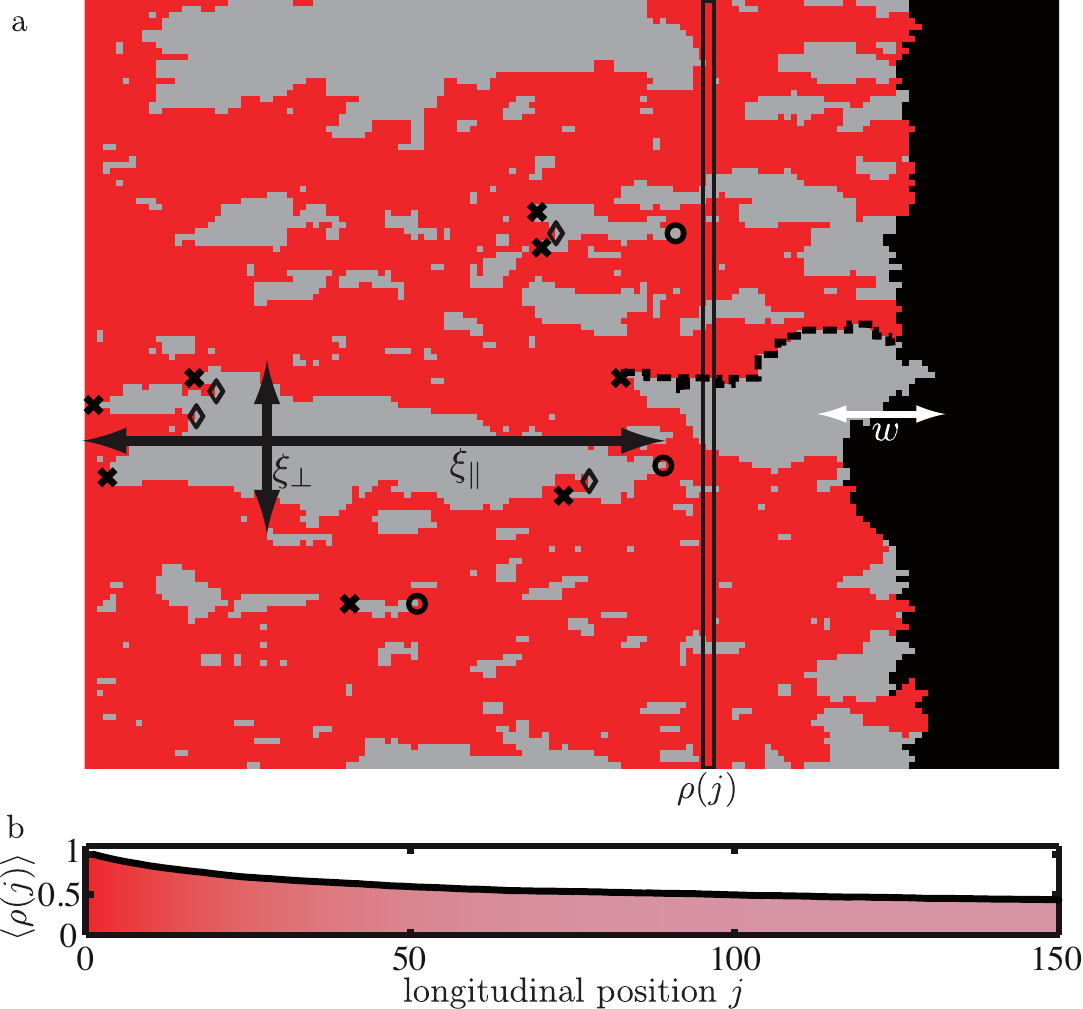}
\caption{\label{fig:observables}Phenomenology of the model (color online). (a) Morphology. Wild-types are shown in dark gray (red), mutants in light gray. During range expansion mutants are created (examples indicated by crosses, $\times$), which in turn reproduce and thereby create mutant clusters. The lateral boundaries (an example is given in the dashed black line) of mutant clusters are created pairwise and perform biased, super-diffusive random walks. Mutant clusters are either outgrown by the wild-types (examples indicated by circles, $\bigcirc$), or they merge  with other clusters (examples marked with diamonds, $\Diamond$), forming even bigger mutant clusters. In both cases a pair of boundaries annihilates. In the bulk, the presence of mutant clusters is discernible as a density of wild-type sites $\rho(j) < 1$  at longitudinal position $j$. Mutant clusters in the bulk are characterized by their typical longitudinal and transverse extensions, which are identified with the correlation lengths $\xi_\parallel$ and $\xi_\perp$, respectively. The surface roughness of the population front is characterized by the width $w$, defined as the standard deviation of the front position. For this realization on a lattice of length $L = 128$, we used birth rate $b = 0.9$ and mutation probability $p=0.016$.  (b) Averaged observables. By averaging over many independent realizations of the growth process fluctuation effects are suppressed and mean observables can be defined. Shown here is the mean density $\langle \rho(j) \rangle$ as function of longitudinal position $j$, averaged over  500 realizations for the parameters used in (a). For large $j$, the density of wild-type sites settles to a stationary value $\rho_s$, as creation and annihilation of mutant clusters (and boundaries) at the front equilibrate.}
\end{figure}

We now turn to a phenomenological description of the morphology of the evolving colony, as obtained from stochastic simulations; a representative realization of the range expansion process is shown in Fig.~\ref{fig:observables}a. Starting out from an initial line of wild-types the growth front moves forward and, as a result of the stochastic individual birth processes, some parts of the front expand more rapidly than others. This leads to front \emph{roughening} which first appears on length scales of the lattice constant, but as time progresses, the typical size of protrusions and indentations of the front grows both longitudinally and laterally.

As the range expansion proceeds, \emph{mutation events} occur where a wild-type individual gives birth to a mutant, whose direct descendants create a new mutant sector growing between two wild-type domains. Mutant and wild-type domains are separated by domain boundaries. Since the reproduction rates of mutant and wild-type individuals differ in general, the boundary between their respective growth sectors performs a \emph{biased random motion}. While for beneficial mutations, with $b > 1$, sectors consisting of mutants broaden on average, they have a tendency to decrease in size for deleterious mutations where $b < 1$. Fluctuations in the trajectory of the boundary arise mainly for two reasons: On short length scales, they are due to the intrinsic stochasticity of the birth events. On larger scales, roughening of the front drives a super-diffusive meandering of the sector boundaries~\cite{Derrida:1991tv,Saito:1995wl,Hallatschek:2007gv}: As the population locally always expands normal to the front, the roughness of the front is imposed on the trajectories of the sector boundaries.

While the domain boundaries move transversely through the system, they may encounter other boundaries, resulting in mutual annihilation and an ensuing merging of domains. There are two distinct types of coalescence events of domain boundaries, cf.~Fig.~\ref{fig:observables}a: Either boundaries of \emph{different} mutant clusters meet such that they merge and form a larger mutant cluster, or two boundaries of the \emph{same} cluster meet, in which case a mutant cluster loses contact to the growing front and is trapped by wild-types. Note that boundaries are always created and annihilated pairwise, and regions of wild-types and mutants alternate at the front.

The relative frequency of events creating and annihilating sector boundaries determines the ultimate fate of the expanding front. In wild-type dominated regions of the front, mutation events enhance phenotypic heterogeneity and create new boundaries. At the same time, merging of mutant clusters creates homogeneous mutant regions. If the selective advantage of wild-types is too small to trap mutant clusters, coalescence events promote growth of mutant clusters, leading to more uniform front populations. Since we do not allow for back-mutations, the expanding front may end up in an \emph{absorbing state} where it is completely taken over by mutants. For a finite system, this will eventually always happen even for deleterious mutations. The main question to ask then is how the corresponding \emph{fixation time} scales with the system size $L$. 

Inspecting Fig.~\ref{fig:observables} one can discern several morphological features of the expanding population, which we discuss phenomenologically now, and analyze quantitatively in the following sections. We discriminate properties of the front and of the bulk of the population. An important bulk observable is the \emph{density} of wild-type sites,  
\begin{eqnarray}
\rho(j) := \frac1L \sum_i \delta_{s_{i,j},1} \; ,
\end{eqnarray}
at longitudinal position $j$. As indicated in Fig.~\ref{fig:observables}b, the ensemble averaged density decays,  for $L \to \infty$,  towards some stationary value $\rho_s$, which serves as an order parameter. A value of $\rho_s = 0$ corresponds to the absorbing state where all individuals at the front are mutants. Any finite value indicates phenotypic heterogeneity with both mutants and wild-types present at the front of the expanding population. Mutant clusters are, in general, anisotropic and one has to distinguish between their extension parallel and perpendicular to the preferred direction of the range expansion: The corresponding longitudinal and transverse \emph{correlation lengths} are denoted by $\xi_\parallel$ and $\xi_\perp$, respectively. The front of the population is characterized by its average speed and its roughness. A good measure for the latter is the \emph{width} $w$, defined as the standard deviation of the front's position,
\numparts
\begin{eqnarray}
\label{eq:width}
w(L,t) &:= \left(\frac1L\sum_{i=1}^L \left[ h(i,t) - \bar h(t)\right]^2 \right)^\frac12 \; ,
\end{eqnarray}
from its average value,
\begin{eqnarray}
\bar h(t) &:= \frac1L\sum_{i=1}^L h(i,t)\; .
\end{eqnarray}
\endnumparts
Here $h(i,t)$ is the local position of the front, defined as the largest $j$ for which $s_{i,j} \neq 0$.

\begin{figure}[tb]
\centering
\includegraphics{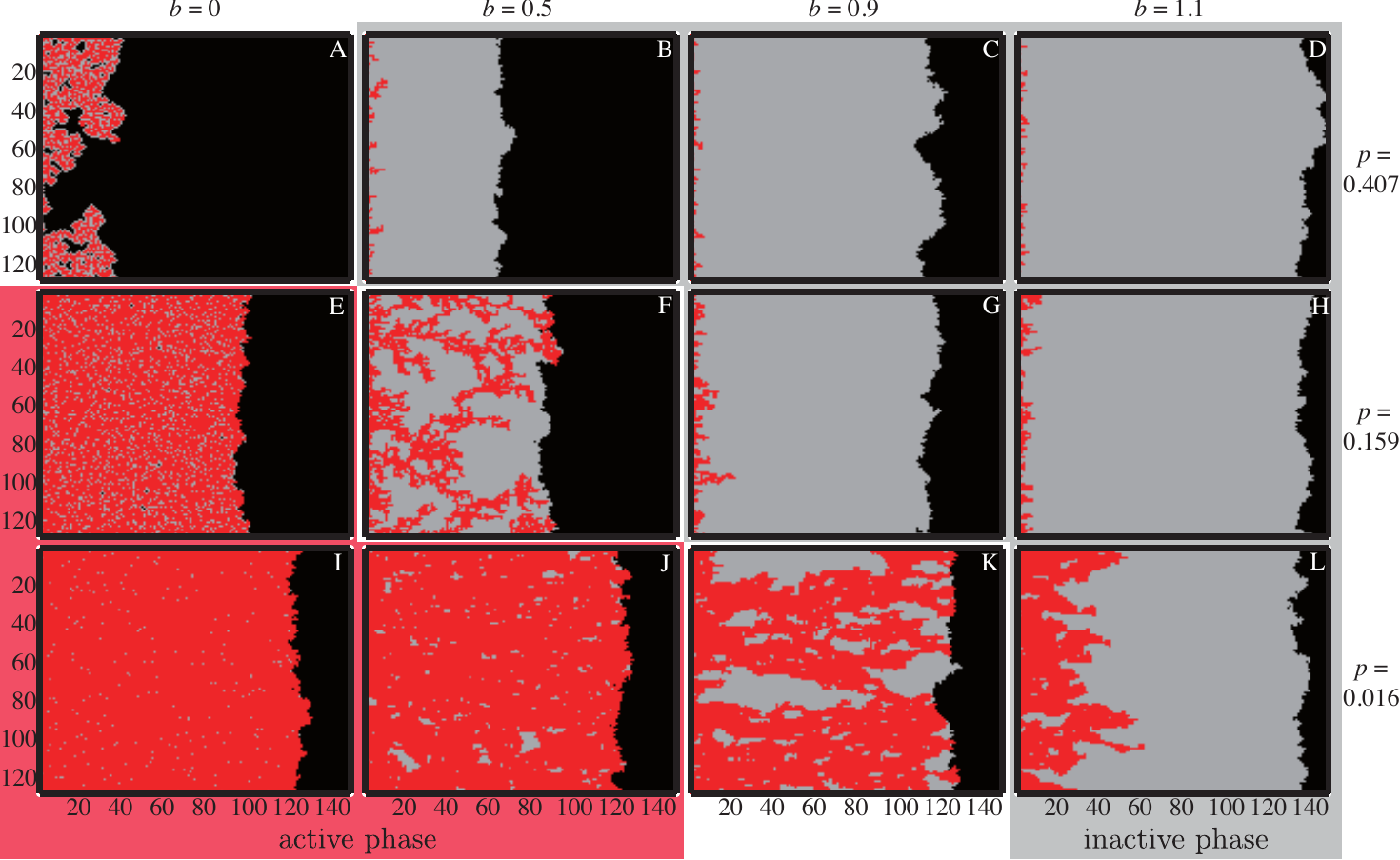}
\caption{\label{fig:examples} Parameter study of the Eden model with mutations (color online). Typical realizations of the model in a finite system ($L=128$) are shown for time $t=100$ and various combinations of mutation rates $p$ and relative birth rates $b$. Wild-types are shown in dark gray (red), mutants in light gray, and empty sites in black. For beneficial mutations (cf.~panels D, H, and L) mutant clusters expand quasi-deterministically. Together with panels B, C, and G these are examples for the inactive phase. Here the wild-type is lost in the bulk and the absorbing state, with only mutants at the front, is reached very fast, as mutants merge into extended clusters. 
As opposed to this, the active phase (cf.~panels E, I, and J) is characterized by long lasting heterogeneous fronts with both mutants and wild-types present. Here, the average spatial distance between distinct mutation events is larger than the extension of the created mutant clusters, which repeatedly lose contact to the front. A non-equilibrium phase transition separates the two phases, where low relative birth rates $b$ and high mutation probabilities $p$ balance and rich self-affine patterns evolve (cf.~panels A, F and K). Going from panel J to panel B at constant birth rate $b = 0.5$, the system changes from a heterogeneous front of wild-types and mutants to a homogeneous all-mutant front. The transition is characterized by diverging length scales (mutant clusters), vanishing wild-type density, and enhanced width of the population front, as discussed in the main text.}
\end{figure}

\subsection{Phase Behavior: Active and Inactive Phase}

The morphology of the expanding population, as discussed above, depends on the values of the mutation probability $p$ and the relative reproduction rate $b$ of the mutant individuals, cf.~Fig.~\ref{fig:examples}. One can clearly identify two distinct phases: In what we call the \emph{active phase} the population front is composed of both wild-types and mutants in a heterogeneous mixture. Here mutants are continuously created by mutation events and the ensuing mutant sectors are subsequently lost again by a coalescence process where sector boundaries meet and the domain of mutants looses contact to the front. This typically happens in a parameter regime where mutations are deleterious and rare. In contrast, if mutations become more frequent and/or their reproduction rate becomes larger, mutant clusters have a significant probability to merge and/or to sweep through the system, and thereby completely take over the population front. This is termed the \emph{inactive phase} since the absorbing state, with only mutants at the front, cannot be left anymore.  

For all beneficial mutations, where $b > 1$, we are well in the inactive phase, independent of the probability $p$ at which mutations appear. Here, a sector created from a single mutation has a finite opening angle and hence grows laterally on average, as observed in experiments~\cite{Shapiro:1995id,Hallatschek:2010kb}. Therefore, already a single mutant can sweep through the population and take over the population front, as the boundaries of the respective growth sector have, on average, a transverse velocity directed outwards~\cite{Saito:1995wl,Murray:2007ux,Hallatschek:2010kb}. This fixation process is further accelerated by the merging of neighboring mutant sectors, cf.~Fig.~\ref{fig:examples} panel L. 

For deleterious mutations, where $b<1$, there is an antagonism between merging of different mutant clusters, creating large mutant domains, and closing of individual mutant clusters, leading to an enlarged fraction of wild-types at the front. Since mutations go from wild-type to mutant only, phenotypic heterogeneity is maintained as long as there are some wild-type individuals left at the front. 

An isolated sector of mutants can survive only for a finite time interval due to stochastic effects. These enable mutants to ``surf'' population waves in expanding populations~\cite{Edmonds:2004il,Excoffier:2008hw,Travis:2007iy,Hallatschek:2008jd}: If mutations appear at the front of an expanding population they have a twofold advantage compared to mutations in spatially homogeneous settings. First, the front can be seen as a perpetual population bottleneck, where demographic fluctuations are enhanced, which in turn reduces the effect of selection. Second, offspring of mutations at the front can spread into unoccupied territory, where there is less competition. By this ``founder effect'' mutants form clusters at the front that can reach much higher frequencies and evade extinction much longer than would be expected in spatially homogeneous settings. However, in the long run an isolated sector tends to loose contact to the front as it is outgrown by the wild-type as a result of the lower birth rate of mutant individuals, cf. panels E, I, and J of Fig.~\ref{fig:examples}. The phenomenology changes qualitatively when mutation events become more frequent. Then, nearby mutant sectors can merge and thereby counteract the loss of mutant sectors by the coalescence of sector boundaries \cite{Hallatschek:2010kb}, cf.~panels B, C, and G, of Fig.~\ref{fig:examples}. The founder effect promotes growth of individual mutations, but persistence of mutants is guaranteed by merging of domains. As a consequence of this antagonism one expects a phase boundary $p_c(b)$ between the passive and the active phase. 
Hallatschek \emph{et al.}~considered mutations appearing at a flat front and found that for deleterious mutations there exists a critical mutation rate where mutants take over. We here consolidate this interesting finding and incorporate the roughness of the front. We in detail analyze the properties of the transition from a heterogeneous to an all mutant front in Section~\ref{sec:crit_behav}.

\subsection{\label{sec:phasediagram}Phase Diagram}

Since our model explicitly excludes back-mutations, the absorbing state, for which the wild-type strain has lost contact with the population front and only mutant individuals are present, cannot be left. For finite systems, $L < \infty$, this absorbing state is eventually always reached, as even for deleterious mutations mutant clusters of arbitrary size can appear through rare fluctuations. As noted above, the key quantity to analyze is the average time for this to happen as a function of mutation probability $p$, relative birth rate $b$ and system size $L$. This \emph{mean fixation time} $t_f(p,b,L)$ is defined as the mean time $t$ when the number of wild-type individuals with empty neighbors, $N_{wt}$, becomes zero for the first time. 

\begin{figure}[tb]
\centering
\includegraphics{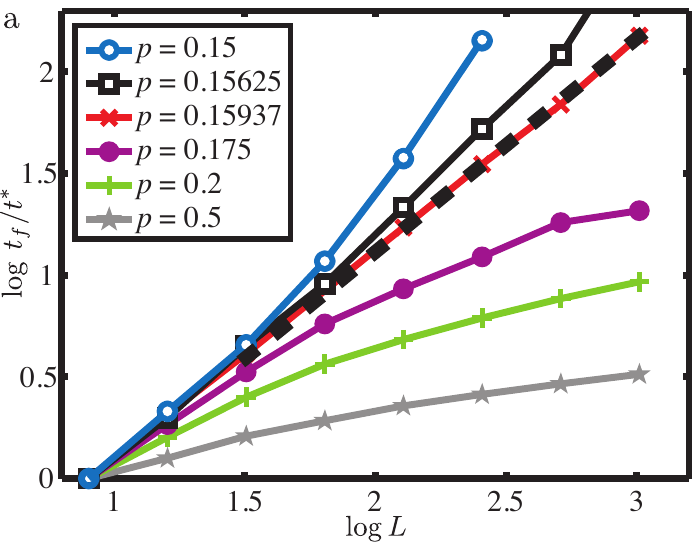} \hspace{0.5cm}
\includegraphics{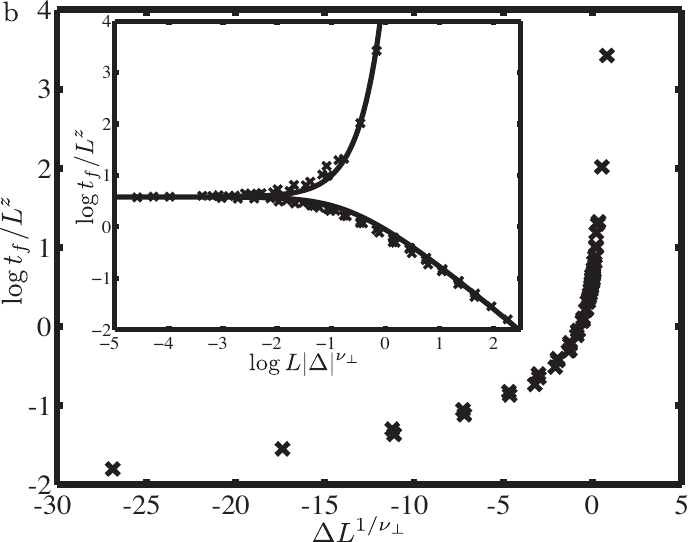}
\caption{\label{fig:ftvsl} Fixation times $t_f$ as a function of the system size $L$ (color online). (a) The fixation time $t_f$ is the mean time until the absorbing state, where only mutants are at the expanding front, is reached. Asymptotically the fixation time grows with the system size $L$. Three cases, as discussed in the text and Fig.~\ref{fig:examples}, can be distinguished. Fast fixation, $t_f \sim \ln L$, exponentially slow fixation, $t_f \sim \exp cL$, and the marginal case $t_f \sim L^z$, where $z = 1.05 \pm 0.05$ (dash black line). Shown here are fixation times for $b = 0.5$, but for different birth rates the discrimination of three distinct regimes holds. For comparability, fixation times have been normalized by $t^* := t_f(L=8)$. (b) Same data as in (a), but rescaled with system size $L$ and distance to the phase transition $\Delta$, according to finite size scaling as detailed in Section \ref{sec:bulk}. All data collapses onto a master curve as can be inferred from the semi-logarithmic plot. The inset is a double logarithmic plot of the same data, which depicts the characteristic scaling above and below criticality, $t_f = L^z \tau^\pm(L/\Delta^{-\nu_\perp})$. As a guide to the eye we included best fits (black lines) to these functions given by $\tau^+(x) = 3.8\exp(10.5x)/(1+x)$ and $\tau^-(x) = 0.26\ln(1+10.5x)/x$.}
\end{figure}

The mean fixation time generally diverges with growing system size, $L \to \infty$. The results from our stochastic simulations, shown in Fig.~\ref{fig:ftvsl}, allow to distinguish three generic cases: (i) For large mutation probability $p$ or beneficial mutations $b \ge 1$ we find $t_f \sim \ln L$. This is the same result as for well-mixed populations~\cite{Frey:2010iz,Cremer:2009kh}. We take this asymptotic law as a hallmark for the inactive phase. Note that for isolated beneficial mutations (neglecting merging of mutant clusters) one finds that the extinction time scales linearly in the system size~\cite{Hallatschek:2010kb}. Hence, the logarithmic law arises from the merging events. 
(ii) In the active phase we have $t_f\sim \exp (cL)$, with some constant $c$. (iii) The phase boundary $p_c(b)$ between the active and the inactive phase is characterized by power law behavior of the mean extinction time:  $t_f\sim L^z$ with a dynamical exponent $z$. This signature is well known from previous studies of phase transitions to absorbing states~\cite{henkel2009book1,Odor:2004wm} (see Section \ref{sec:crit_behav}). From our numerical data we estimate $z = 1.05 \pm 0.05$. 

\begin{figure}[tb!]
\centering
\includegraphics{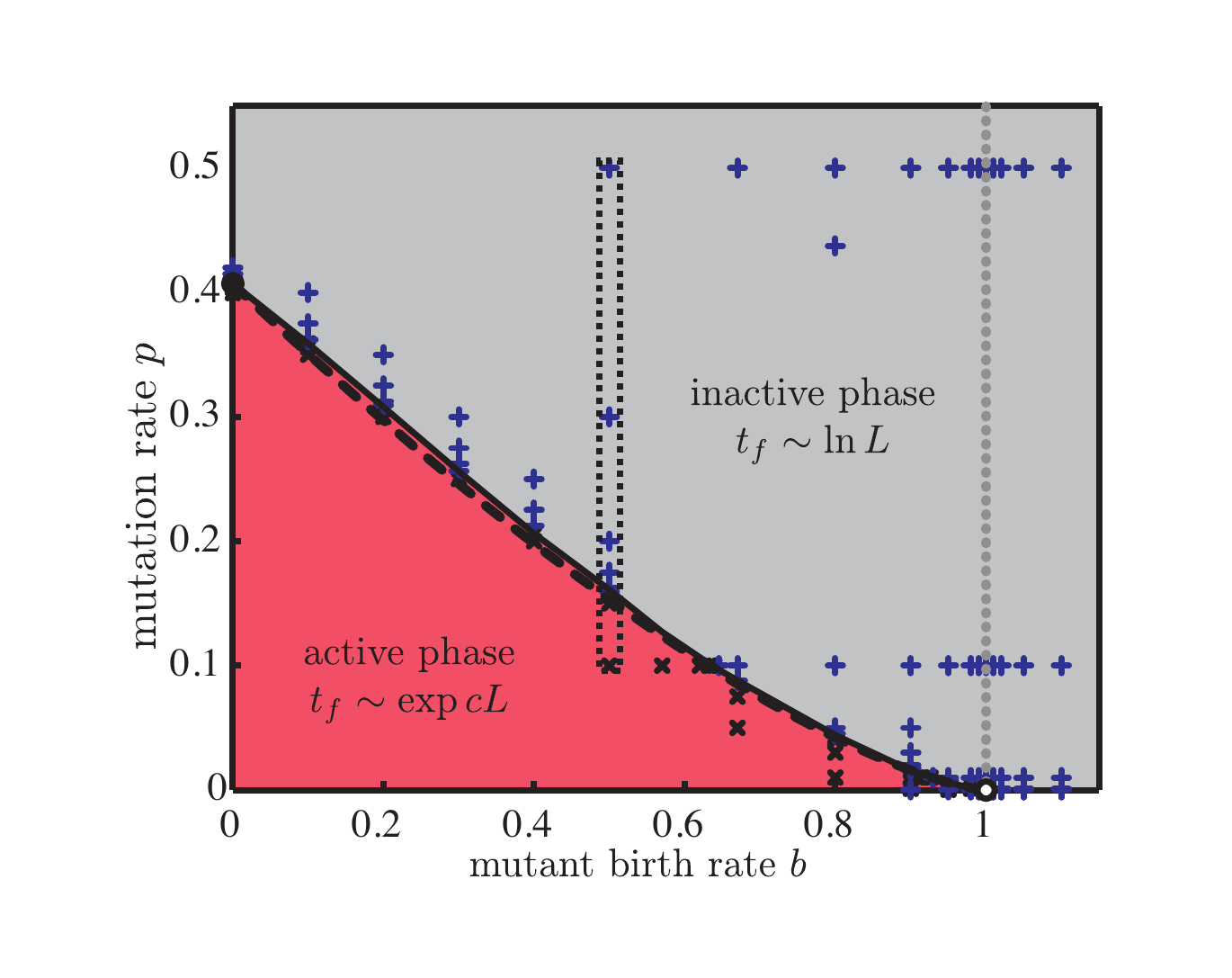}
\caption{\label{fig:phase_diagram}Phase diagram of the Eden process with mutations (color online). The transition from the inactive phase (fast fixation, light gray) to the active phase (exponentially slow fixation, dark gray [red]) is a non-equilibrium phase transition (solid black line). Simulations can be categorized by the slope in a double-logarithmic $t_F$ vs.~$L$ plot for $L \gg 1$: Plus signs (+, blue) denote inactive phase behavior with slope $< 1$; crosses ($\times$, black) denote active phase behavior with slope $> 1$. Along the gray dotted line $b=1$ mutations are neutral. The point $(p = 0, b=1)$ (black open circle) is not part of the transition line, as for $p=0$ no mutants can appear and we recover the original Eden model. The phase transition line terminates on the axis $b=0$, at critical mutation probability $p^*$ (full black circle). This point corresponds to the critical density of isotropic site percolation, see main text. The mean field transition line (black dashed line), given by Eq.~(\ref{eq:mf_trans_line}), is in excellent accordance with our numerical results. To characterize the properties of the phase transition we explore how physical observables depend on the distance to the phase transition $\Delta := p_c -p$. To this end, simulations were carried out for $b=0.5$ (without loss of generality), indicated by a black dotted box, for which we found $p_c = 0.159 \pm 0.001$.}
\end{figure}

Since the phase transition from the active phase to the inactive phase is accompanied by a qualitative change in the $L$-dependence of the mean fixation time $t_f$ from logarithmic to exponential behavior, we may use it to map out the phase diagram, see Fig.~\ref{fig:phase_diagram}.

We find a phase transition line $p_c(b)$, which ends at the points $(b,p)=(0,p^*)$ and $(b,p)=(1,0)$. For $b=0$ mutants do not reproduce at all and each mutant is created by a reproducing wild-type individual in an independent event with probability $p$. For low $p$ the clusters have a typical size of one site and are obstacles around which the wild-types grow. For larger $p$ more extended clusters are formed by mutation events that happen on neighboring sites. For the wild-types this corresponds to an Eden process on a \emph{site percolation} network, where wild-type individuals are identified with occupied sites (present with probability $1 - p$) and mutant individuals with empty sites (present with probability $p$). Thus $(b,p)=(0,p^*)$ is a multi-critical point of the transition line, below which the wild-type strain, for $L \to \infty$, keeps growing on the infinite percolating cluster. This implies 
that $p^* = 1 - p_{c}^\textrm{\scriptsize{isp}} \approx 0.407$, where $p_{c}^\textrm{\scriptsize{isp}}$ is the percolation threshold of isotropic site percolation on a 2d square lattice~\cite{stauffer1994book}. 

For non-vanishing birth rates, $b > 0$, a newly formed mutant cluster can grow to some significant size. As a result clusters originating from more distant mutation events can merge before loosing contact to the front. The phase transition line marks the set of mutation rates $p_c (b)$, where, for a given birth rate of mutants $b$, the typical spatial distance between two mutation events becomes comparable to the typical extension of individual mutant clusters. Thus, as we move to larger and larger birth rates, the critical mutation probability $p_c$ declines. 

For $b \to 1$ mutations become neutral, i.e., selection pressure vanishes and the random motion of sector boundaries is not biased anymore. Individual clusters can grow unlimited by fluctuations, and consequently the smallest number of mutations suffices for the population to be in the inactive phase. If $p=0$ no mutants are created and we recover the classical one-species \emph{Eden model} composed of growing wild-types~\cite{Eden:1960vd}. Thus, at $(b,p)=(1,0)$ the front is composed of only wild-type individuals and the absorbing state is never reached. Our numerical data suggests that $p_c$ approaches this point in a cusp-like singularity.  

For $b > 1$  mutations are beneficial, which means mutations and selection are not antagonists anymore, and hence a phase transition is absent. Here, the mutants take over the system quasi-deterministically for all finite values of $p$.

To rationalize our findings for the phase diagram we consider a phenomenological mean field theory for the dynamics of the density of the wild-type strain at the front of the expanding population, $n = N_{wt}/(N_{wt}+N_{mut})$. On average, wild-type individuals are lost by mutation at a rate $p$, and gained or lost through natural selection depending on the relative growth rates of the wild-type and mutant individuals. The effect of natural selection is proportional to $n(1-n)$, which vanishes if the front is composed of one strain only. This leads to the following rate equation
\begin{eqnarray}
\dot n = -p n + s(b)n(1-n) \, , 
\label{eq:mfmodel}
\end{eqnarray}
where $s(b)$ is an \emph{effective selection strength}. The functional form of $s(b)$ can be determined phenomenologically: The rate equation~(\ref{eq:mfmodel}) has the fixed points $n_1=0$ and $n_2 = 1 - p/s(b)$. In the active phase, $n_1=0$ is unstable, while $n_2$ is stable. At the phase transition the two fixed points merge and interchange their stability in a transcritical bifurcation. Solving $n_1 = n_2$ gives the mean field transition line $p_{c}^\textrm{\scriptsize{mf}}(b) = s(b)$. For $b \to 0$ the problem reduces to isotropic percolation and hence $s(0) = p^*$.  
Certainly the selection coefficient $s(b)$ has to vanish if mutants reproduce as fast as wild-types, $s(1)=0$, which gives us the other end of the transition line. Close to it the transition line can be approximated by a power law, $p_{c}^\textrm{\scriptsize{mf}}(b \to 1) \sim (1-b)^\mu$, reflecting the cusp singularity we found in the simulations. The simplest ansatz for the mean field transition line, which fulfills these constraints is
\begin{eqnarray}
	 p_{c}^\textrm{\scriptsize{mf}}(b) = s(b) =p^* (1-b)^\mu \; . \label{eq:mf_trans_line}
\end{eqnarray}
A fit to our numerical derived phase transition line gives $\mu = 1.41 \pm 0.03$. The phenomenological mean-field transition line, depicted in Fig.~\ref{fig:phase_diagram}, is in very good agreement with our numerical results.

\section{\label{sec:crit_behav}Critical Behavior}

In  this section, we in-depth investigate the properties of the absorbing state phase transition from the active into the inactive phase. Without loss of generality we focus on a fixed value for the birth rate of mutants, $b = 0.5$, for which we found the critical mutation probability to be $p_c(b=0.5) = 0.159 \pm 0.001$. The qualitative behavior of all observables stays the same along the transition line. Close to the special points $b=0$ and $b=1$ crossover effects become more pronounced, which we do not examine here. 

\subsection{\label{sec:bulk}Bulk Properties}

We first address bulk properties of the system, that is observables measured sufficiently far away from the rough growth front. Since we are dealing with a phase transition to an absorbing state, we expect to find four independent critical exponents~\cite{Hinrichsen:2000wg, Hinrichsen:2006cn, Odor:2004wm, henkel2009book1}. A  common choice of observables is: The stationary density $\rho_s := \rho(j \to \infty)$ of active sites (wild-types); the survival probability 
$P_s$ (Survival in this context means the wild-type having contact to the growth front.) of a single active site (wild-type) in a front of inactive sites (mutants); the longitudinal correlation length $\xi_\parallel$, and the transverse correlation length $\xi_\perp$. For $L \to \infty$, all these observables diverge like power laws in the control parameter $\Delta := p_c - p$ in the vicinity of the phase transition. The respective critical exponents are defined through
\numparts
\begin{eqnarray}
\rho_s &\sim \Delta^\beta \; \; \textrm{ for }\Delta \ge 0 , \label{eq: beta}\\
P_s &\sim \Delta^{\beta'} \; \textrm{ for }\Delta \ge 0 , \label{eq: betadash}\\
\xi_\parallel &\sim \left| \Delta \right|^{-\nu_\parallel} \; , \label{eq: xipar}\\
\xi_\perp &\sim \left| \Delta \right|^{-\nu_\perp} \label{eq: xiperp}\; . 
\end{eqnarray}
\endnumparts
The above  observables are understood as ensemble averages. The stationary density $\rho_s$ and the survival probability $P_s$ are 0 in the inactive phase, $\Delta < 0$, since in this case any heterogeneous composition of the front is unstable.

For finite systems, $L < \infty$, both the stationary density $\rho_s$ and the survival probability $P_s$ are identical to $0$ for the inactive \emph{and} the active phase. Though it may be extremely rare, it is always possible -- even in the active phase -- that large enough mutant clusters appear and finally lead to a front consisting of mutants only, which is the absorbing state. Hence, $\rho_s$ and $P_s$ are not particularly useful observables for $L < \infty$, and one instead has to examine the time-dependent density $\rho(\Delta,t,L)$ and survival probability $P(\Delta,t,L)$, and also time-dependent correlation lengths $\xi_\parallel(\Delta,t,L)$ and $\xi_\perp(\Delta,t,L)$. 

\begin{figure}[tb]
\centering
\includegraphics{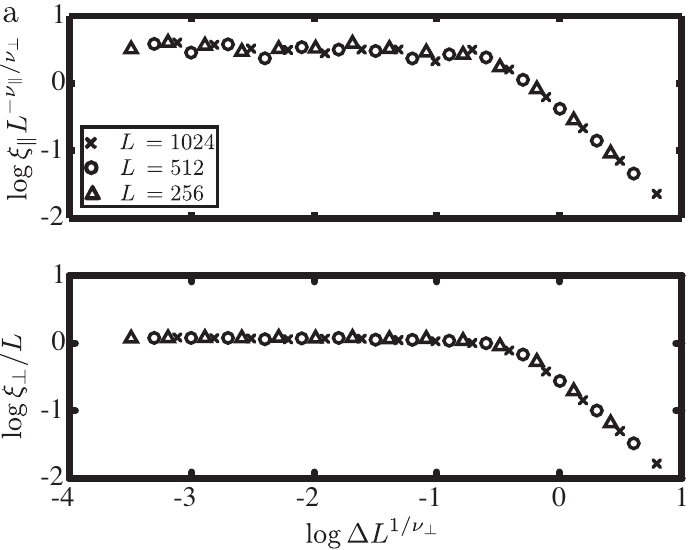} \hspace{0.5cm}
\includegraphics{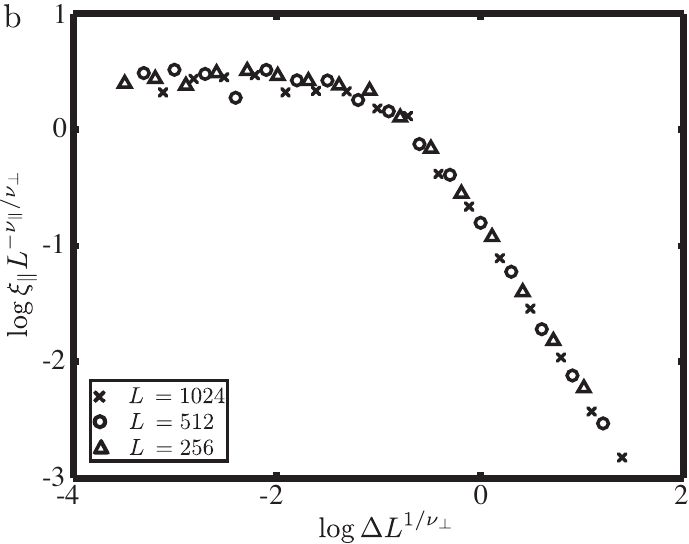}
\caption{\label{fig:clengths} Rescaled correlation lengths from simulations started with a line of wild-types as initial condition for different lattice sizes. (a) Rescaled correlation lengths from cluster-size distributions in the active phase. Longitudinal correlation length $\xi_\parallel$ (upper plot) and transverse correlation length $\xi_\perp$ (lower plot) have been computed from mutant cluster masses $m$ and cluster extensions $\ell_\parallel$ and $\ell_\perp$ (see text). Correlation lengths depend on both $L$ and distance to the phase transition $\Delta$. Taking $\nu_\perp = \nu_\parallel = 1.6 \pm 0.1$ all data collapse to master curves under rescaling. (b) Rescaled longitudinal correlation length as found from decay of wild-type density in the inactive phase. The wild-type density $\rho$ decays exponentially, with a decay length proportional to the longitudinal correlation length. Again all data collapse to a master curve for $\nu_\perp = \nu_\parallel = 1.6 \pm 0.1$.}
\end{figure}

We measured the critical exponents, defined in Eqs.~(\ref{eq: beta}--\ref{eq: xiperp}), using different methods. 
Starting with a line of wild-types in the active phase we let the system evolve, disregarding realizations where the absorbing state was reached. This allows us to use stationary observables, instead of more involved dynamical ones. From each realization we extracted for each mutant cluster i) the longitudinal extension $\ell_\parallel$, ii) the transverse extension $\ell_\perp$ and, iii) the number of cluster sites or mass $m$. In analogy to percolation theory~\cite{stauffer1994book}, the correlations lengths are then calculated by 
\begin{eqnarray}
\xi ^2 _{\#} (\Delta,L)= \frac{2\sum_k m_k \ell_{\#,k}^2}{\sum_k m_k} \, ; \; \# \in \{\parallel,\perp\} \;,
\end{eqnarray}
where $k$ runs over all observed clusters. The computed correlation lengths depend on both the distance to the phase transition $\Delta$ and the system size $L$. However, close to the critical point all macroscopic observables are invariant under scaling transformations of the form~\cite{Hinrichsen:2000wg, Hinrichsen:2006cn, Odor:2004wm, henkel2009book1}
\numparts
\begin{eqnarray}
\Delta & \to c \Delta \; , \label{eq:scaling_trans1}\\
x_\parallel & \to c^{-\nu_\parallel} x_\parallel \; , \label{eq:scaling_trans2}\\
x_\perp & \to c^{-\nu_\perp}  x_\perp \; , \label{eq:scaling_trans3}\\
\rho & \to c^\beta \rho \; , \label{eq:scaling_trans4}\\
P & \to c^{\beta'} P \; \label{eq:scaling_trans5}.
\end{eqnarray}
\endnumparts
where $c$ is some positive rescaling factor.  This implies for the correlation lengths the following scaling forms, 
\begin{eqnarray}
\xi_{\#}(\Delta,L) = L^{\nu_{\#}/\nu_\perp} f_{\#}(\Delta L^{1/\nu_\perp}) \, ,
\end{eqnarray}
where the $f_{\#}(x)$ are universal scaling functions with the asymptotic behavior
\begin{eqnarray}
f_{\#}(x) \sim
\left\{
\begin{array}{ll}
1 & x \ll 1 \\
x^{-\nu_{\#}} & x \gg 1 \, .
\end{array}
\right.
\end{eqnarray}
Indeed, after rescaling, our simulation data for both correlation lengths $\xi_{\#}(\Delta,L)$ collapse onto master curves for all $L$ and $\Delta$, if we take
\begin{eqnarray}
\nu_\perp =  \nu_\parallel = 1.6 \pm 0.1\; , \label{eq:xis}
\end{eqnarray}
see Fig.~\ref{fig:clengths}a. Note that this means that close to the transition the longitudinal and transverse correlation length show the same scaling behavior, $\xi_\perp \sim \xi_\parallel$, at least within the accuracy of our simulations. Simply put, a fluctuation of twice the size takes twice as long to decay, which explains that close to the phase transition fixation times are proportional to the system size, $t_f \sim L$, which we found in Section \ref{sec:phasediagram}. 

With the critical exponents $\nu_\perp$ and $\nu_\parallel$ at hand, rescaling of the fixation time $t_f(L,\Delta)$ is easily achieved as well. Note that as a time-like observable $t_f$ scales like a longitudinal distance. Again employing the phenomenological scaling theory for absorbing state phase transitions, Eqs.~(\ref{eq:scaling_trans1}--\ref{eq:scaling_trans5}), we obtain
\begin{eqnarray}
t_f(L,\Delta) = L^z \tau^\pm(\Delta L^{1/\nu_\perp}) \, ,  
\end{eqnarray}
where $z= \nu_\parallel / \nu_\perp$ is the dynamical critical exponent. The overall scaling function can be split up into two distinct parts, $\tau^+$ and $\tau^-$, which exhibit characteristic logarithmic and exponential behavior for $\Delta > 0$ and $\Delta < 0$, respectively. Using this scaling, all data for the fixation times collapse, as shown in Fig.~\ref{fig:ftvsl}b.

In the inactive phase, the density $\rho(\Delta,j,L)$ decays exponentially with $j$ and the absorbing state, with $\rho = 0$, is reached fast. 
The decay length is proportional to $\xi_\parallel$ which can easily be found by fitting an exponential function to the measured density profile. Applying finite size scaling, we again find data collapse for critical exponents $\nu_\perp = \nu_\parallel = 1.6 
\pm 0.1$, see Fig.~\ref{fig:clengths}b.

Further evidence for these values of the critical exponents is obtained by calculating the (active state) correlation functions
\numparts
\begin{eqnarray}
\Gamma_\parallel(r) &= \langle s_{i,j} s_{i,j+r}\rangle - \langle s_{i,j}\rangle\langle s_{i,j+r}\rangle\; , \\
\Gamma_\perp(r) &= \langle s_{i,j} s_{i+r,j}\rangle - \langle s_{i,j}\rangle\langle s_{i+r,j}\rangle\; ,
\end{eqnarray}
\endnumparts
in the bulk. Averages are with respect to all lattice indices $i,j$ and independent realizations. The correlation functions are expected to behave as 
\begin{eqnarray}
\Gamma_{\#}(r) \sim r^{-\sigma_{\#}} \exp\left(-\frac{r}{\xi_{\#}}\right), \; \# \in\{\parallel,\perp\} \; , 
\end{eqnarray}
where $\sigma_{\#} = 2 \beta / \nu_{\#}$.
To obtain the correlation lengths we fitted this expression to our data with the fitting parameters $\xi_{\#}$. In plots of the correlation lengths, we again find good collapse of data for $\nu_\perp = \nu_\parallel = 1.6 \pm 0.2$ (not shown).

\begin{figure}[tb]
\centering
\includegraphics{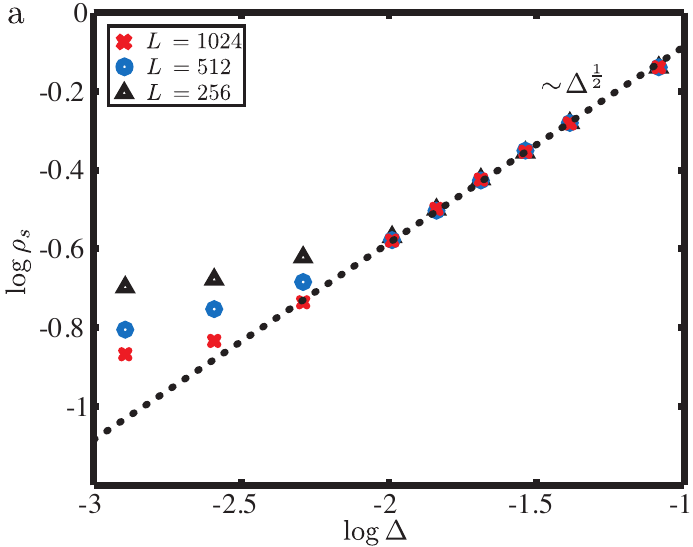}\hspace{0.5cm}
\includegraphics{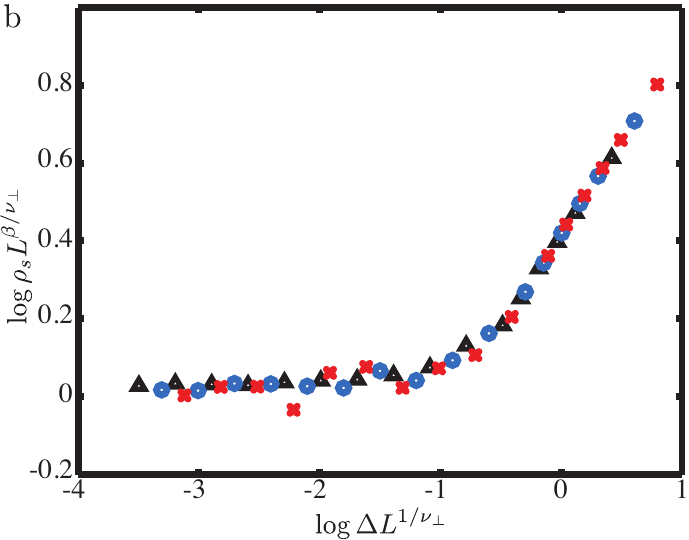}
\caption{\label{fig:density} Wild-type stationary density in the active phase (color online). (a) Simulations for different lattice sizes started with a line of wild-types as initial condition in the active phase. We find a power law with critical exponent $\beta = 0.50 \pm 0.02$ that, for small $\Delta$, holds better in larger systems, a typical finite size effect. (b) Same data as in (a) after finite size scaling. We find collapse of all data for critical exponents $\beta = 0.5$  and $\nu_\perp = 1.6$.}
\end{figure}

Next, we consider the stationary density $\rho_s(\Delta)$, which equals the fraction of wild-type sites for large $j$, that is sufficiently far away from the initial line for the density to relax to its stationary value. As shown in  Fig.~\ref{fig:density}a, a double-logarithmic plot asymptotically gives power law behavior with the exponent 
\begin{eqnarray}
\beta = 0.50 \pm 0.02 \; . \label{eq:beta}
\end{eqnarray}
Deviations from the asymptotic power law are well described by the finite-size scaling form 
\begin{eqnarray}
\rho_s(\Delta,L) = L^{-\beta/\nu_\perp} g(\Delta L^{1/\nu_\perp}) \, ,
\end{eqnarray}
where $g(x)$ is a universal scaling function, cf.~Fig.~\ref{fig:density}b. Two characteristic regimes can be distinguished
\begin{eqnarray}
g(x) \sim
\left\{
\begin{array}{ll}
1 & x \ll 1  \\
x^\beta & x \gg 1 \, ,
\end{array}
\right.
\end{eqnarray}
corresponding to the scaling law $\rho_s \sim \Delta^\beta$, for a given $L$ and sufficiently far from criticality, and the scaling law, $\rho_s \sim L^{-\beta/\nu_\perp}$, for a given distance $\Delta$ from the critical point and system sizes smaller than the transverse correlation length, $L \ll \xi_\perp$.

\begin{figure}[tb]
\centering
\includegraphics{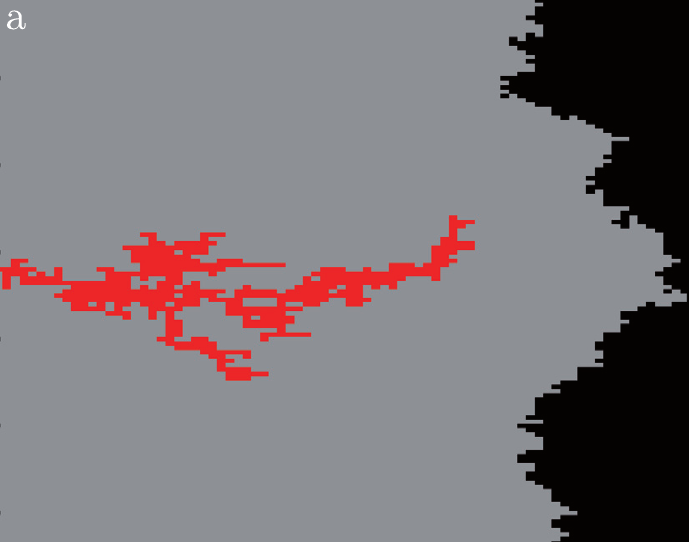} \hspace{0.5cm}
\includegraphics{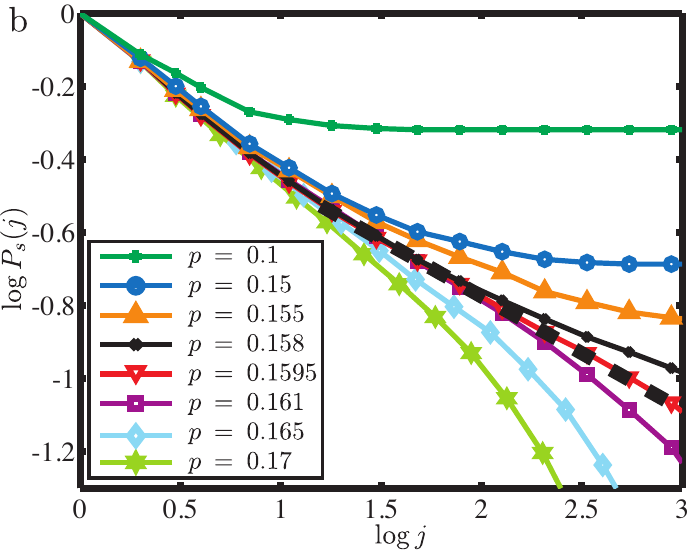}
\caption{\label{fig:ssic} Single wild-type initial condition and related observables (color online). (a) Typical realization for single-seed initial conditions. Wild-types are shown in dark gray (red) mutants in light gray. By averaging over many realizations of the process the survival probability of wild-types $P(j)$ and the number of wild-types $N(j)$ can be measured as a function of longitudinal position $j$. For this realization on a lattice of length $L = 128$ we used birth rate $b = 0.5$ and mutation probability $p=0.15$.
(b) Survival probability as a function of time. Simulations for different mutation rates $p$ started with a single wild-type site as initial condition. For $p_c \approx 0.159$ we find a power law for $P_s(j)$ with critical exponent $\beta'/\nu_\parallel = -0.32 \pm 0.2$, indicated by the dashed line. Birth rate $b = 0.5$ and lattice size $L = 1024$ are the same for all shown data.}
\end{figure}

To measure $\beta'$, the exponent associated with the survival probability when starting from a single seed, $P_s$, different initial conditions must be used. Instead of a line composed of only wild-types, a single wild-type is placed in a line of mutants. A typical realization for these initial conditions is shown in Fig.~\ref{fig:ssic}a.
From an ensemble of realizations the survival probability $P_s(j)$ and the mean number of wild-types $N(j )$, both as functions of distance $j$ to the initial line, can be extracted. After transients have decayed and sufficiently far away from the front (where there might exist empty lattice sites), we find power laws for both quantities when close to the phase transition. For $P_s$ the exponent is given by $-\beta'/\nu_\parallel$ while for $N$ it is denoted by $\theta$. Fig.~\ref{fig:ssic}b shows results for $P_s(j)$, from where we obtain $\beta'/\nu_\parallel = -0.32 \pm 0.2$. Using our result for $\nu_\parallel$ we find
\begin{eqnarray}
\beta'  = 0.51 \pm 0.07 \;. \label{eq:betadash}
\end{eqnarray}
Similarly, from the simulation data for $N(j)$ we obtain 
\begin{eqnarray}
\theta = 0.32 \pm 0.02\; .
\label{eq:exponent_theta}
\end{eqnarray}
$\theta$ is related to the other critical exponents by the generalized hyperscaling relation~\cite{Mendes:1994wj} in $d=1$ dimension
\begin{eqnarray}
\theta = \frac{\nu_\perp - \beta -\beta'}{\nu_\parallel} \; . \label{eq:hsr}
\end{eqnarray}
Using our previous results for the critical exponents we find good agreement with the above result, Eq.~(\ref{eq:exponent_theta}), within the error margin. The critical exponents for the phase transition to an absorbing state are summarized in Table \ref{tab:crit_exponents}.
\begin{table}[tbh]
\caption{\label{tab:crit_exponents}Critical exponents for the non-equilibrium phase transition to an absorbing state of the two-species Eden model with mutations and selection.}
\begin{indented}
\item[]\begin{tabular}{@{}cccccc}
\br
$\beta$ & $\beta'$ & $\nu_\parallel$ & $\nu_\perp$ & $z$ & $\theta$\\
\mr
$0.50 \pm 0.02$ & $0.51 \pm 0.07 $ & $1.6 \pm 0.1$ & $1.6 \pm 0.1$ & $1.05 \pm 0.05$ & $0.32 \pm 0.02$\\
\br
\end{tabular}
\end{indented}
\end{table}

For systems with infinitely many absorbing states, like ours, it is known that critical exponents, especially those related to single seed initial conditions (in our case $\beta'$ and $\theta$), can vary significantly with the configuration away from the seed~\cite{Jensen:1993wg,Munoz:1996tc}. The correct exponents, which satisfy the hyperscaling relation of Eq.~(\ref{eq:hsr}), are obtained if the configuration away from the seed is a typical configuration of the absorbing state. In our case this amounts to a rough front. To check for this dependence of the critical exponents, we initialized single wild-type seeds at arbitrary front positions of all-mutant Eden models grown to saturation. Data extraction works similar, but due to roughness and the non-uniqueness of the growth direction the initial seed is not necessarily the one with the smallest $j$-value, which complicates the evaluation slightly. These simulations take more time since independent saturated fronts have to be generated for every realization, which makes it harder to obtain data with high precision. However, we find $\beta'/\nu_\parallel = -0.33 \pm 0.03$ and $\theta= 0.31 \pm 0.04$ (data not shown), which indicates that our system is robust with respect to the initial configuration. We are therefore confident, that the critical exponents of the phase transition, see Table \ref{tab:crit_exponents}, are indeed correct up to the given precision.

To check if the DP universality class is recovered if roughness is neglected, we also considered a simplified model with synchronous update rules. Here, an individual at site $(i,j)$ has two possible parents, at sites $(i-1,j-1)$ and $(i,j-1)$. Depending on their states, we define the probabilities $P(s_{i,j}|s_{i-1,j-1},s_{i,j-1})$ for the type of offspring $s_{i,j}$, incorporating mutations and selection: If both parents are mutants, the offspring is inevitable mutant as well; if both parents are wild-types the offspring is wild-type if no mutation happens; if one parent is wild-type while the other is a mutant, the mutant reproduces with rate $b/(1+b)$ and the wild-type otherwise. In the latter case a mutation can happen, resulting in a mutant offspring. All events are summarized by the probabilities
\numparts
\begin{eqnarray}
&P(1|-1,-1) = 0 \; ,\\
&P(1|1,-1) = P(1|-1,1) := p_1= \frac{1-p}{1+b}\; ,\\
&P(1|1,1) := p_2 = 1-p \; ,
\end{eqnarray}
\endnumparts
and $P(-1|.,.) = 1- P(1|.,.)$. It is easily seen that this corresponds to the Domany-Kinzel cellular automaton~\cite{Domany:1984ud}, with probabilities $p_1$ and $p_2$. Indeed, for $b=0.5$ we find the critical mutation probability for the flat model $p_c^\textrm{\scriptsize{flat}} \approx 0.077$, corresponding to $p_1 \approx 0.62$ and $p_2 \approx 0.93$. This lies well on the transition line of the Domany-Kinzel automaton~\cite{Lubeck:2006bq}. As one would expect, the critical exponents of the flat model are that of the DP universality class (data not shown). These conclusions are further consolidated by earlier studies for a two-species ballistic deposition model with kinetics that allows surface growth only at sites where one type of particles has an exposed position on top of the incidence or neighboring column~\cite{Reis:2002jn}.  These particular growth rules imply that the surface roughness does not affect the bulk dynamics such that it can be mapped onto a one-dimensional contact process~\cite{Harris:1974ue} which is in the DP universality class~\cite{Hinrichsen:2000wg,Hinrichsen:2006cn,Odor:2004wm,henkel2009book1}. In conclusion, the interplay between surface roughening and domain dynamics in our two-species Eden model is clearly responsible for the deviation of its critical behavior from that of the DP universality class.

\subsection{\label{sec:surface}Front Properties}

We now turn to characterizing the front of the expanding population, i.e.~to surface roughness properties of the model. As argued in Section \ref{sec:phenomenology}, the presence of mutations changes the roughening behavior of the growth front qualitatively: Close to the phase transition large mutant clusters form extended parts of the leading front and since mutants and wild-types reproduce with different rates, the front's roughness  is strongly enhanced. In the case of detrimental mutations, $b<1$, mutant-dominated regions trail behind compared to the average front; see for example Fig.~\ref{fig:examples}, panel F. 

The width is the key observable in the analysis of kinetic surface roughening processes. These processes can be organized into universality classes, which are characterized by symmetry properties and the dimensionality of the process under consideration~\cite{godreche1991solids,HalpinHealy:1995wb,Barabasi:1995p4091}. Each growth model's universality class has a unique set of two exponents: The growth exponent $\gamma$ and roughness exponent $\alpha$, defined through
\begin{eqnarray}
w(L,t) \sim 
\left\{
\begin{array}{ll}
t^\gamma & t \ll t_{\times} \\
L^\alpha & t \gg t_{\times} \; ,
\end{array}
\right. \label{eq:width_scale}
\end{eqnarray}
with the crossover time between these two regimes scaling as $t_{\times} \sim L^{\tilde z}$, where ${\tilde z}=\alpha/\gamma$ is the dynamic roughening exponent.

\begin{figure}[t!]
\centering
\includegraphics{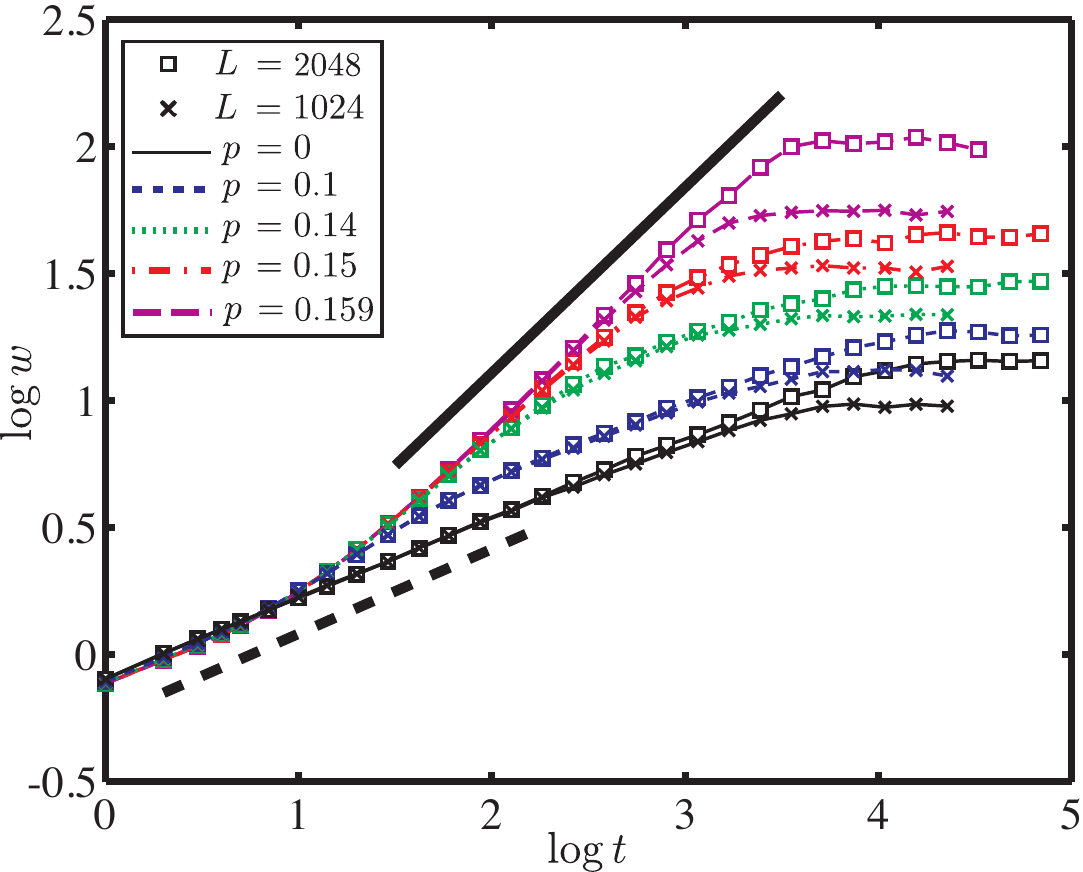}
\caption{\label{fig:wvst} Width of the growth front as a function of time in the active phase (color online). The temporal evolution of the width $w(t)$ of the growth front depends on both the system size $L$ and the distance to the phase transition $\Delta$. $w$ initially grows Eden-like, $\gamma = 1/3$ (bold dashed line). Close to the phase transition (i.e.~at small $\Delta$) we find a crossover to a growth exponent $\gamma' = 0.73 \pm 0.03$ (bold straight line), as a result of different growth rates of mutants and wild-types. For intermediate times two scenarios can be distinguished, see main text for details: (i) Either the surface width saturates at a finite length-dependent value as a result of finite system size, or (ii) mutant clusters at the front saturate at a size smaller than $L$. In the latter case Eden growth is recovered on large length and time scales. As the width saturates we recover the Eden roughness exponent $\alpha = 1/2$ but with a $p$-dependent amplitude of the saturation width.  In the former case, enhanced roughening proceeds until saturation. The saturated width then scales like $L^{\alpha'}$, with a different roughness exponent $\alpha' = 0.91 \pm 0.01$.
All data shown are for $b=0.5$, but results hold for different birth rates $b$ as well.}
\end{figure}

In the limit of vanishing mutation rate, $p = 0$, our model reduces to the Eden model, which falls into the well-known KPZ universality class~\cite{Kardar:1986vl}, with exponents $\gamma = 1/3$ and $\alpha = 1/2$. For $p > 0$, the time evolution of the width depends on whether the system is in the active phase or in the inactive phase. In the inactive phase we find a transiently enhanced width as compared to the Eden model and a subsequent return to it. We attribute this to mutants outcompeting wild-types rapidly, as is typical for the inactive phase, followed by Eden growth of the resulting mutant front (data not shown). As shortly discussed in Section \ref{sec:pandp}, the active phase is characterized by a heterogeneous front. In finite systems, this is the case until a mutant cluster covers the whole system size $L$, which then takes the system into the absorbing state with subsequent return to Eden-like growth. Therefore, the roughness characteristics of the heterogeneous front in the active phase can only be analyzed if the absorbing state has not yet been entered. Hence, for the following analysis, realizations which reach the absorbing state are stopped, and simulation data are used only up to this point in time.

We now discuss the different regimes and scenarios of surface roughening in the active phase. Fig.~\ref{fig:wvst} displays our numerical results for a range of mutation rates and two different systems sizes. Since we start with an initial condition where the lattice only contains wild-types, the initial surface roughening is that of a single species, i.e.~Eden growth: $ w \sim t^\gamma$ with $ \gamma = 1/3 $. As time passes, mutants are created, leading to growing mutant clusters. Their extension in growth direction evolves like the correlation length $\xi_\parallel(\Delta,t,L)$, which initially grows proportional to $t$, as follows from scaling. Since these clusters grow slower than wild-type clusters this leads to additional surface roughening. As for small $t$ both contributions to roughening are independent of $\Delta$ and $L$, the width gets dominated by differential expansion velocities at a crossover time $t_{\times,1} = \mathcal{O}$(1). 

As a result of the ensuing \emph{strong coupling} between domain dynamics and surface roughness, the interface width then grows more rapidly than in the Eden model. We observe a regime with an altered growth exponent $\gamma' = 0.73 \pm 0.03$, which becomes more extended upon approaching the phase transition to the inactive phase, $\Delta \to 0$. There are two distinct scenarios by which this enhanced roughening regime may end: (i) Either the surface width saturates at a finite length-dependent value as a result of finite system size, or (ii) mutant clusters at the front saturate and their typical size reaches the asymptotic extension $\xi_\perp^\infty \sim \Delta^{-\nu_\perp} \leq L$.\footnote{A third scenario where the typical lateral extension of mutant clusters reaches the extension of the system $\xi_\perp \sim L$ is also possible but excluded by our sampling method.}

\begin{figure}[tb]
\centering
\includegraphics{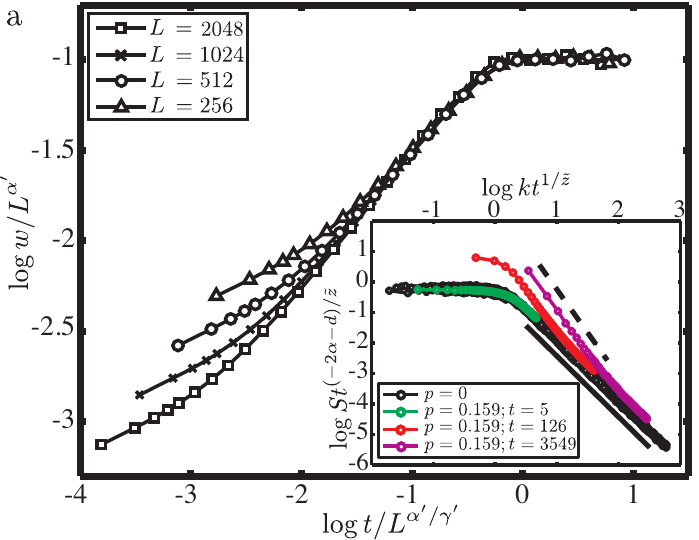} \hspace{0.5cm}
\includegraphics{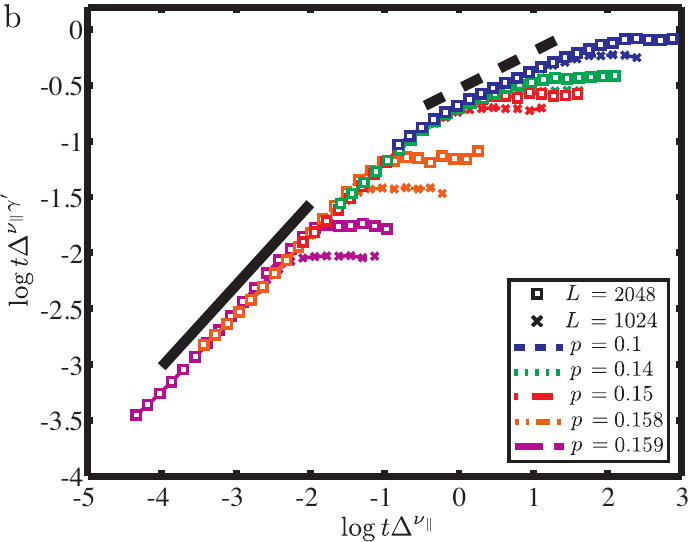}
\caption{\label{fig:wvst_scaled} Scaling plots for surface roughening in the Eden model with mutation (color online): 
(a) \emph{Critical scaling}. Rescaled width versus time at criticality ($p = 0.159$) for a series of system sizes as indicated in the graph. Neglecting initial transients, all data collapses for $\alpha' = 0.91\pm 0.01$ and $\tilde z' = \alpha'/\gamma' = 1.25\pm 0.05$. Both growth exponent $\gamma'$ and roughening exponent $\alpha'$ are markedly different from their KPZ counterparts. The inset shows the rescaled structure factor  $S(k,t)$ for system size $L = 512$. For the Eden model all data collapses, and from the black solid line, proportional to $k^{-2}$, we recover $\alpha = 1/2$ from Eq.~(\ref{eq:strucfactscaling}). For the critical system, $\Delta = 0$, we find an extended scaling regime with $S \sim k^{-2.85 \pm0.05}$ (black dashed line) corresponding to $\alpha' = 0.92 \pm 0.03$. (b) \emph{Crossover scaling}. Neglecting initial transients, data can be rescaled with respect to the intrinsic crossover time $t_{\times,2} \sim \Delta^{-\nu_\parallel}$, where the mutant cluster extension $\xi_\perp$ saturates at $\Delta^{-\nu_\perp}$. If the system size $L$ is much smaller than  $\xi_\perp^\infty$, the front width grows to saturation with the new growth exponent $\gamma' = 0.73 \pm 0.03$ (black solid line), whereas for $\xi_\perp^\infty \ll L$ the saturation width is reached after return to Eden-like roughening with $\gamma = 1/3$ (black dashed line). Saturation, due to finite size, happens where curves deviate from the common envelope and become horizontal.}
\end{figure}

The first case requires that one is sufficiently close to the critical point such that the cluster size $\xi_\perp$ does not saturate at $\Delta^{-\nu_\perp} \leq L$, but mutant clusters continue to grow during the whole roughening process. Then, our simulations show that there is a direct crossover from enhanced surface roughening, $w \sim t^{\gamma'}$, to saturation, $w \sim L^{\alpha'}$ with a roughness exponent $\alpha' = 0.91 \pm 0.01$. The corresponding scaling form close to criticality reads
\begin{equation}
w_c (L,t) = L^{\gamma'} \hat w_c \bigl( {t/L^{\tilde z'}} \bigr)
\end{equation}
with the dynamic critical exponent for surface roughening $\tilde z'={\alpha'}/{\gamma'} = 1.25\pm 0.05$, cf.~Fig.~\ref{fig:wvst_scaled}a. (Note that deviations from the critical scaling behavior are all due to initial transient Eden growth: all systems were started with an initial condition of wild-type individuals only.) A roughening exponent $\alpha'$ close to unity indicates a very jagged surface. For models with large roughness exponents, especially for super-roughening with exponents larger than $1$ (see, e.g.,~\cite{DasSarma:1994ur,Lopez:1997ui}), it is known that they may exhibit anomalous scaling~\cite{Plischke:1993bq,Schroeder:1993wk}, in the sense that the roughness exponent as determined from $w(L,t \to \infty)$ may only be an effective exponent. The actual exponent can be obtained from the structure factor or power spectrum~\cite{Schmittbuhl:1995kz,Siegert:1996gu,Barabasi:1995p4091}
\begin{equation}
S(k,t) := \langle h(k,t)h(-k,t) \rangle \; ,
\end{equation}
where $h(k,t)$ is the Fourier transform of the height fluctuations.
From Eq.~(\ref{eq:width_scale}) it follows that the structure factor of the Eden model scales as
\begin{equation}
S(k,t) = t^{(d+2\alpha)/\tilde z}s(kt^{1/\tilde z}) \; ,
\end{equation}
with the scaling function
\begin{eqnarray}
s(x) \sim
\left\{
\begin{array}{ll}
1 & x \ll 1  \\
x^{-d-2\alpha} & x \gg 1 \, .
\end{array}
\right. \label{eq:strucfactscaling}
\end{eqnarray}
The rescaled structure factor of our model is shown as the inset of Fig.~\ref{fig:wvst_scaled}a. For the trivial case $p=0$ we observe data collapse as expected for the Eden model. At criticality $\Delta = 0$ ($p=0.159$), the scaling function for the power spectrum changes with time due to the growth dynamics of mutant clusters, i.e. the time dependence of the cluster size $\xi_\perp (t)$. For asymptotically large times, when a single mutant cluster spans the system, a broad regime emerges with a  roughness exponent $\alpha' = 0.92\pm 0.03$ corroborating our results obtained from analyzing the interface width.

This clearly shows that, for $\Delta \to 0$, the Eden model with mutations and selection is no longer in the KPZ universality class, but exhibits different asymptotic roughening behavior. We attribute this behavior to the strong coupling between critical domain growth and surface roughening.

Moving further away from the critical point, typical domains may saturate at their asymptotic value $\Delta^{-\nu_\perp} \leq L$ within a crossover time $t_{\times,2}\sim \Delta^{-\nu_\parallel}$. Then, on length scales larger than $\xi_\perp^\infty$, roughening proceeds, but slower than before, and one recovers Eden-like roughening with $w \sim t^\gamma$. This regime is most pronounced if $L\gg \xi_\perp^\infty$, and finally saturates in a constant surface width with $w  \sim L^\alpha$. The crossover between critical roughening and return to Eden roughening is determined by a time scale inherent to the system which only depends on the distance to the critical point but not on the system size. We may, therefore, rescale time by the crossover time, $t_{\times,2} = \Delta^{-\nu_\parallel}$, and rescale width $w$ by the corresponding value $w(t_{\times,2}) = t_{\times,2}^{\gamma'} = \Delta^{-\gamma' \nu_\parallel}$, cf. Fig.~\ref{fig:wvst_scaled}b. All curves collapse to a crossover scaling function, from which they only deviate due to finite size effects.

\section{\label{sec:discussion}Conclusion and Outlook}

A generalization of the Eden model to a two-species model, with uni-directional mutations and selection due to different reproduction rates, has interesting properties from the viewpoints of both dynamical phase transitions as well as kinetic surface roughening. We find that surface roughness has a marked effect on the critical properties of the absorbing state phase transition. While reference models~\cite{Hallatschek:2010kb}, which keep the expanding front flat, exhibit DP critical behavior, the exponents of our generalized Eden model strongly deviate from it. In turn, the mutation-selection process in the bacterial colony induces a increased surface roughness with exponents distinct from that of the KPZ universality class. 

For our model the critical exponents of the longitudinal and transverse correlation length, $\nu_\parallel$ and $\nu_\perp$, are identical within the error margin. As the aspect ratio $\xi_\parallel/\xi_\perp$ can still be different from unity, this does not make our system isotropic. However, it indicates that there is no longer a rigorous distinction between a time-like longitudinal and a space-like transverse direction as for DP. To some degree, this might have been expected from the microscopic update rules of the Eden model, which have no preferential direction of growth. Indeed, for vanishing birth rate of mutants, the system can be mapped to \emph{isotropic} percolation, and it looses any preferential growth direction.  One may also argue that front roughening in concert with locally normal growth of the interface leads to ``mixing'' of transversal and longitudinal directions, and thereby promotes the same scaling behavior of both directions. It remains to be seen how this result can be obtained from a renormalization group calculation. 

The critical exponents summarized in Table \ref{tab:crit_exponents} are significantly different from those of the DP universality class, which are observed for a broad range of systems exhibiting phase transitions to absorbing states~\cite{Hinrichsen:2000wg, Hinrichsen:2006cn, Odor:2004wm, henkel2009book1}. The stability of the directed percolation universality class is reflected in the DP hypothesis, a conjecture formulated by Janssen~\cite{Janssen:1981wr} and Grassberger~\cite{Grassberger:1982ub}. It asserts that models characterized by a positive one-component order parameter, which exhibit a phase transition to an unique absorbing state, belong to the DP universality class, provided that interactions are short-ranged and no additional symmetries or special kinds of disorder or fluctuations are present~\cite{Hinrichsen:2000wg, Hinrichsen:2006cn, Odor:2004wm, henkel2009book1}. Strictly speaking, our model is not in contradiction with the DP hypothesis as it does not fulfill the condition of having a \emph{unique} absorbing state. Instead, there are infinitely many configurations with only mutants at the front. However, even for models with infinitely many absorbing states one often finds DP-class universal behavior~\cite{Jensen:1993wg,Munoz:1996tc}. What distinguishes our model is that growth persists in the fluctuating absorbing state of an all mutant system. Most importantly, there is interplay between surface roughness and bulk properties: Even though we analyzed bulk properties of the system, the dynamics are restricted to the front and, therefore, convey imprints of the rough surface. This might also be a reason why it is difficult to find actual experimental systems that obey DP-like behavior: Any non-flat system may carry signatures of surface roughening in the numerical values of the critical exponents of the corresponding non-equilibrium phase transition.

The influence of surface roughness on domain boundaries, without the additional complication of mutations, selection and of merging clusters, is a rather complex problem on its own. The morphology of the boundary between two distinct but equally fast growing Eden clusters has  previously been analyzed by Derrida and Dickman~\cite{Derrida:1991tv}. Numerically they found that the dynamics of the domain boundary depends on the local curvature of the front. It may perform a subdiffusive or superdiffusive random walk, and even move ballistically. For our model this implies enhanced complexity, because boundaries are preferentially created at protrusions, where wild-types are more prone to be, while merging of domain boundaries usually occurs at indentations of the front. Coarsening of domains in a two-species Eden model is in the focus of work by Saito and M\"uller-Krumbhaar~\cite{Saito:1995wl}, who also considered differential reproduction rates of the two species. As mutations are absent in their model, the faster growing species outgrows the slower one exponentially fast. In the neutral case they show that domain boundaries exhibit super-diffusive scaling and thereby explain the domain coarsening kinetics of their simulations. It would be interesting to find a suitable generalization of their analysis to include the case of irreversible mutations.

There are also growth models where surface roughness and particle configuration at the front mutually affect each other~\cite{Ausloos:1993ts, Ausloos:1996vs, Kotrla:1997vx, Kotrla:1998vf, Drossel:2000vw, Drossel:2003cwa}. A theoretically well-studied system is vapor deposition of binary films containing two different kind of molecules which have a tendency to phase separate~\cite{Drossel:2000vw, Drossel:2003cwa}. Similar to the dynamics of the two-species Eden model, there is an interplay between surface roughening and phase ordering kinetics: the dynamics of the domain boundaries and the surface are coupled through the growth kinetics. One main difference between the models is that for binary films the non-equilibrium roughening process is coupled to the coarsening dynamics of a thermodynamic model (Ising model) exhibiting detailed balance, while in our model it is coupled to the far-from-equilibrium dynamics of an absorbing state phase transition. In the binary film particles of one type attach to domains containing particles of mainly the other type. If this is interpreted as ``mutation", then mutations are bi-directional and symmetric in the binary film. In addition, the growth rules differ in some important aspects. While in our two-species Eden model, the two species have different growth rates, in binary film growth, particle attachment at domain boundaries and within domains differ~\cite{Ausloos:1993ts, Ausloos:1996vs, Kotrla:1997vx, Kotrla:1998vf, Drossel:2000vw, Drossel:2003cwa}. A common finding of these studies on binary films is that phase ordering kinetics increases surface roughness on length scales comparable to domain size; this is phenomenologically similar to our findings but the critical exponents are quite different. The reverse coupling of surface roughness to phase ordering kinetics is subtle and depends on the details of the kinetic growth rules. If growth at domain boundaries is faster than within domains, the surface roughness imprints its scaling properties on the domain boundaries~\cite{Drossel:2003cwa}, as in our case. Else, it appears that the domain boundaries perform random walks or even show non-universal behavior~\cite{Derrida:1991tv, Drossel:2000vw, Drossel:2003cwa}. 

The enhanced roughness, which is induced by differential front velocities, bears some similarities with pinning models~\cite{Barabasi:1995p4091}.  There, inhomogeneities (e.g.~obstructions) of the medium locally reduce the growth speed, similar to mutant clusters in our model. The presence of these heterogeneities changes the scaling of the interface, since it introduces a length scale $\xi_\perp$, up to which one finds enhanced roughening.  It is found that in the pinned phase, many pinning models can be mapped to DP, and the roughness of the surface is dominated by the DP critical exponents~\cite{Buldyrev:1992fe,Tang:1992kr}. In contrast to pinning models, where the disorder is quenched, the mutant clusters in the Eden model with mutation are dynamically generated and strongly correlated with the growth dynamics. We regard the latter as the main reason why the ensuing roughness and growth exponents are not related to DP.

Summarizing, we here argue that for multi-species range expansion, surface roughness and domain dynamics interfere with each other which qualitatively changes both bulk and front properties. If absorbing front-states exist, phase transitions to absorbing states of a new universality class are possible. More research is needed to explore the coupling between surface roughening and evolutionary dynamics of range-expansion scenarios. It would be especially interesting to look at systems where the reproduction rate of an individual depends on the local composition of the front as might be the case in growing biofilms~\cite{Nadell:2009de}. Such systems, where growth depends on a population's composition, have been analyzed in the context of  one dimensional wave propagation~\cite{Hallatschek:2011wf}, the dilemma of cooperation in spatial settings~\cite{Nadell:2010fe}, games with cyclic dominance~\cite{Rulands:2011vm}, and structured populations~\cite{Cremer:2011ut}. A discrete 2$d$ growth model, which incorporates these effects, would be most interesting to analyze with the methods developed in this paper. Moreover, the universality class of our model should be tested for different lattices, and critical exponents should be determined for higher dimensional setups. \\

\ack
Financial support by the Deutsche Forschungsgemeinschaft, Grant FR850/9-1 and RA655/5-1, and the Elite Network of Bavaria (International Doctorate Program NanoBioTechnology) is gratefully acknowledged.

\section*{References}

\bibliography{edenmutmanuskript}

\end{document}